\documentclass[preprint,5p,twocolomn]{elsarticle}

 \usepackage{amsfonts}
\usepackage{amssymb}
 \usepackage{graphicx}
 \usepackage{widetext}
 \usepackage{amsmath}
  \usepackage{caption}
\usepackage{subcaption}
\usepackage{xcolor}
\usepackage{cuted}

\journal{Journal of \LaTeX\ Templates}

\def\be{\begin{equation}}
\def\ee{\end{equation}}
\def\ba{\begin{array}}
\def\ea{\end{array}}
\def\bea{\begin{eqnarray}}
\def\eea{\end{eqnarray}}

\definecolor{orange}{rgb}{0.75,0.25,0.05}

\DeclareMathOperator{\sech}{sech}
\begin{document}

\begin{frontmatter}

\title{{\color{orange}No scale SUGRA SO(10) derived Starobinsky Model of Inflation}}
\author[PRL]{Ila Garg}
\ead{ila@prl.res.in}
\author[PRL]{Subhendra Mohanty}
\ead{mohanty@prl.res.in}

\address[PRL]{Theoretical Physics Division, Physical Research Laboratory, Ahmedabad 380009, India.}

\begin{abstract}
We show that a supersymmetric renormalizable theory based on gauge group
SO(10) and Higgs system {\bf {10 $\oplus$ 210 $\oplus$ 126 $\oplus$ $\overline{\bf 126}$}} with no scale supergravity
can lead to a Starobinsky kind of potential for inflation.
Successful inflation is possible  in the cases where the  potential during inflation  corresponds
to $SU(3)_C \times SU(2)_L \times SU(2)_R \times U(1)_{B-L}$, $SU(5)\times U(1)$
and flipped $SU(5)\times U(1)$ intermediate symmetry with a suitable choice of superpotential parameters.
The reheating in such a scenario can occur via
non perturbative decay of inflaton i.e. through ``preheating''. After the end of reheating, 
when universe cools down, the finite temperature potential can have 
a minimum which corresponds to MSSM.

\end{abstract}
\begin{keyword}
\texttt{Supergravity} \sep Unified Field Theory \sep Inflation \sep Starobinsky Model
\end{keyword}

\end{frontmatter}

\section{Introduction}
The theory of cosmological inflation \cite{starobinsky,guth,Albrecht} not only solves the problems (flatness, horizon etc.)
of standard big bang theory, but also explains the
seed fluctuations  which can grow via gravitational
instability to form the large scale structure of the universe \cite{mukhanov}. There are stringent constraints on
inflationary theories from CMB observations \cite{COBE,wmap,PLANCK5,BICEP2}  and many of the generic models like
the quartic potential and quadratic potential are either ruled out or disfavoured by the
bound on the tensor to scalar ratio which is $r_{0.05}<0.12$ at 95\% CL from joint analysis of BICEP2/Keck array
and Planck data \cite{BICEP2Planck}. Among the generic inflation models which survive the stringent constraint on
$r$ is the $R^2$ inflation model of Starobinsky \cite{starobinsky} which predicts $n_s-1= -2/N$ and $r=12/N^2 \sim 0.002-0.004$.
The theoretical motivation for the Starobinsky model is provided in \cite{ENO} where
it has been shown that the Starobinsky potential for inflation can be derived from supergravity (SUGRA) with a no-scale
\cite{CFKN,EKN,LN} K\"{a}hler potential and a Wess Zumino superpotential
with specific couplings. Supergravity models of inflation based on the Jordan frame
supergravity \cite{ Einhorn:2009bh,Ferrara:2010in,Ferrara:2010yw}
and D-term superpotential \cite{Buchmuller:2013zfa} also give inflationary potential which
are identical to the Starobinsky potential at large field values.
The natural choice for the inflaton in supergravity models are the Higgs fields of the
grand unified theories. A no-scale SUGRA model of inflation based on the
SU(5) GUT using the {\bf 24}, {\bf 5} and $\overline{\textbf{5}}$ Higgs in the superpotential
has been constructed \cite{Ellis:2014dxa}. The SU(5) symmetry breaks
to MSSM with the appropriate choice of $vev$ for the {\bf 24} and a D-flat linear combination
of $H_u$ and $H_d$ of MSSM acts as the inflaton \cite{Ellis:2014dxa}.

In the present work we study inflation in a renormalizable grand unified theory based on
the SO(10) gauge group with no scale SUGRA. Inflation in the
context of SUSY SO(10) has been studied earlier in \cite{ Kyae:2005vg,Fukuyama:2008dv,Antusch:2010va,Aulakh:2012st,Cacciapaglia:2013tga}
with the SO(10) invariant superpotential with the minimal K\"{a}hler potential which gives
polynomial potentials of inflation.
 In this paper we show that  a renormalizable Wess-Zumino superpotential of SO(10) GUT along with no-scale K\"{a}hler potential
can give us Starobinsky kind of inflationary potential with specific choice of superpotential parameters.
 The Higgs supermultiplets we consider are {\bf10}, {\bf210}, {\bf126} ($\overline{{\bf126}}$).
 Among these, the ${\bf210}$ and ${\bf 126}$ ($\overline{{\bf126}}$) are responsible for breaking of
SO(10) symmetry down to MSSM. The ${\bf 210}$ supermultiplet alone can give different intermediate
symmetries \cite{abmsv} depending upon which of its MSSM singlet field takes a $vev$.
Then ${\bf 126}$ ($\overline{{\bf126}}$) breaks this intermediate symmetry to MSSM.
We find that successful inflationary potential can be achieved in the case of $SU(3)_C \times SU(2)_L \times SU(2)_R \times U(1)_{B-L}$,
$SU(5)\times U(1)$ and flipped $SU(5)\times U(1)$ symmetry. The other possible intermediate symmetries of Pati-Salam
  ($SU(4)_C \times SU(2)_L \times SU(2)_R$) or $ SU(3)_C \times SU(2)_L \times U(1)_R \times U(1)_{B-L}$ 
  gauge groups do not give phenomenologically correct inflationary potentials.

At the end of inflation, the reheating can occur via non perturbative decay of
inflaton to bosons of the intermediate scale model. After the end of reheating, 
when universe cools down, the finite temperature potential can have 
a minimum which corresponds to MSSM and the universe rolls down to this minimum at temperature $<<$ $T_R$ (reheat temperature).

 \section{Inflation in SO(10) with no scale SUGRA}
The minimal supersymmetric grand unified theory based on SO(10) gauge group
\cite{abmsv,aulmoh,ckn,babmoh,bmsv} has \textbf{10}($H_i$), \textbf{210}($\Phi_{ijkl}$)
and \textbf{126}($\Sigma_{ijklm}$)($\overline{\textbf{126}}$($\overline{\Sigma}_{ijklm}$)) Higgs supermultiplets. The
representations $H_i$ is 1 index real, $\Sigma_{ijklm}$ is complex (5 index, totally-antisymmetric, self
dual) and $\Phi_{ijkl}$ is 4 index  totally-antisymmetric tensor. Here $i$, $j$, $k$, $l$, $m$ = 1, 2...10 run over the vector
representation of SO(10).
 The  renormalizable superpotential for the above mentioned fields is given by,
  \bea W&=&\frac{m_{\Phi}}{4!} \Phi^2+\frac{\lambda}{4!} \Phi^3 + \frac{m_{\Sigma}}{5!}
  \Sigma \overline{\Sigma}+\frac{\eta}{4!} \Phi \Sigma \overline\Sigma+m_{H} H^2\nonumber\\&&
  +\frac{1}{4!} \Phi H (\gamma \Sigma+\bar \gamma \overline\Sigma)\,.\eea
 The no-scale form of K\"{a}hler potential is taken to be,
 \be K=-3\ln(T+T^*-\frac{1}{3}(\frac{1}{4!}\Phi^{\dag}\Phi+ \frac{1}{5!}\Sigma^{\dag}\Sigma+
 \frac{1}{5!}\overline{\Sigma}^{\dag}\overline{\Sigma}+H^{\dag}H))\,.\ee
Here $T$ is the single modulus field arising due to string compactification and we are taking $M_{P}$ = 1.

The \textbf{10} and $\overline{\textbf{126}}$ are required for Yukawa terms to
give masses to the fermions while \textbf{126}($\overline{\textbf{126}}$)
 breaks the SO(10) gauge symmetry to MSSM together with $\textbf{210}$-plet.
 However to have a intermediate symmetry rather than MSSM, the \textbf{210}-plet Higgs is sufficient.
 It can lead to various possible intermediate symmetries depending on which components of the  \textbf{210}-plet take $vevs$.
  The decomposition of Higgs supermultiplets required for SO(10) symmetry breaking in terms of Pati-Salam gauge group
  ($SU(4)_C \times SU(2)_L \times SU(2)_R$) is given by  \cite{aulgir},
\bea210 &= (15,1,1)+(1,1,1)+(15,1,3)+(15,3,1)\nonumber\\&+(6,2,2)+(10,2,2)+(\bar{10},2,2)\nonumber\\
  126 &=(\bar{10},1,3)+(10,3,1)+(6,1,1)+(15,2,2)\nonumber\\
 \overline{126} &=(\bar{10},3,1)+(10,1,3)+(6,1,1)+(15,2,2)\nonumber\,.\\
  \eea
 The field components which will not break the MSSM symmetry are allowed to take $vevs$. In this case they are \cite{bmsv},
 \bea p&=&<\Phi(1,1,1)>,\, a= <\Phi(15,1,1)>,\nonumber\\
 \omega&=& <\Phi(15,1,3)>,\,\sigma = <\Sigma(\bar{10},3,1)>, \nonumber\\
 \bar{\sigma}&=&<\bar\Sigma(10,3,1)>\,.\eea
  The Superpotential in terms of these $vevs$ is,
  \bea W &=& m (p^2+3 a^2+6 \omega^2)+2 \lambda(a^3+3p \omega^2+6 a \omega^2)\nonumber\\&&+
  m_{\Sigma}\sigma\bar\sigma+\eta\sigma\bar\sigma(p+3a-6\omega)\label{SP}\,.\eea
The vanishing of D-terms gives the condition $|\sigma|=|\bar \sigma|$ \cite{bmsv}.
The symmetry breaking path of SO(10) is,
  \be SO(10) \xrightarrow{210} \text{Intermediate~~symmetry} \xrightarrow{126} MSSM \nonumber\,.\ee
For the first step symmetry breaking one can set $|\sigma|$ = $|\bar \sigma|$ = 0.
Then the possible intermediate symmetries with \textbf{210} only are \cite{bmsv},
 \begin{enumerate}
 \item If $a$ $\neq$ 0 and $p$ = $\omega$ = 0, it gives $SU(3)_C \times SU(2)_L \times SU(2)_R \times U(1)_{B-L}$ symmetry.
  \item If $p$ $\neq$ 0 and $a$ = $\omega$ = 0, this results in $SU(4)_C \times SU(2)_L \times SU(2)_R$ symmetry.
 \item If $\omega$ $\neq$ 0 and $p$ = $a$ = 0, it gives $ SU(3)_C \times SU(2)_L \times U(1)_R \times U(1)_{B-L}$ symmetry.
  \item If $p$ = $a$ = -$\omega$ $\neq$ 0, this  has $ SU(5) \times U(1)$ symmetry.
  \item If $p$ = $a$ = $\omega$ $\neq$ 0,  $ SU(5) \times U(1)$ symmetry but with flipped assignments for particles.
   \end{enumerate}
The superpotential in terms of $vevs$ of {\textbf {210}} is given by,
\be W = m (p^2+3 a^2+6 \omega^2)+2 \lambda(a^3+3p \omega^2+6 a \omega^2)\ee
Here $m$ = $m_{\Phi}$. Similarly no-scale K\"{a}hler potential is,
\be K= -3\ln(T+T^*-\frac{1}{3}(|p|^2+3|a|^2+6|\omega|^2))\,.\ee
The F-term potential has the following form,
  \be V= e^{G} \biggr[\frac{\partial G}{\partial \phi^i} K^i_{j^*}\frac{\partial G}{\partial \phi_{j^*}}-3\biggr]
 \label{potential}\ee
 Where
 \be G=K+\ln W+\ln W^*\,.\ee
The kinetic term is given as $K_{i}^{j^*} \partial{\phi^i} \partial{\phi_{j^*}}$.
Here $i$ runs over different fields $T,p,a$ and $\omega$.
 $K^i_{j^*}$ is the inverse of K\"{a}hler metric $K^{j^*}_i$ given by,
\be K^{j^*}_i= \frac{1}{\Gamma^2}\begin{pmatrix} 3 & -p^* & -3 a^* & -6 \omega^* \\
-p& \Gamma+\frac{1}{3}|p|^2 & a^* p & 2 \omega^*p\\
-3a & ap^* & 3\Gamma+3|a|^2 &  6 a \omega^* \\
-6 \omega & 2 \omega p^* & 6 a^* \omega & 6\Gamma+12|\omega|^2 \end{pmatrix}\ee
Where $\Gamma=T+T^*-\frac{1}{3}(|p|^2+3|a|^2+6|\omega|^2)$.
After simplifying, the potential given by Eq.(\ref{potential}) has the following form,
\be V= \frac{1}{\Gamma^2} \biggr| \frac{\partial W}{\partial \phi_i}\biggr|^2 \label{PF}\,.\ee
We assume that the non-perturbative Planck scale dynamics \cite{Ellis:2014dxa,ENO,Cicoli:2013rwa}  fixes the values of $T=T^*=\frac{1}{2}$.
After fixing the $vev$ for $T$ the kinetic
terms of $T$ can be neglected.
We study all possible cases of intermediate symmetries mentioned earlier for
inflationary conditions in SO(10) with no-scale SUGRA. For simplicity we assume our fields to be real.\\

\begin{figure}[t]

\begin{subfigure}{.5\textwidth}
  \includegraphics[width=.8\linewidth]{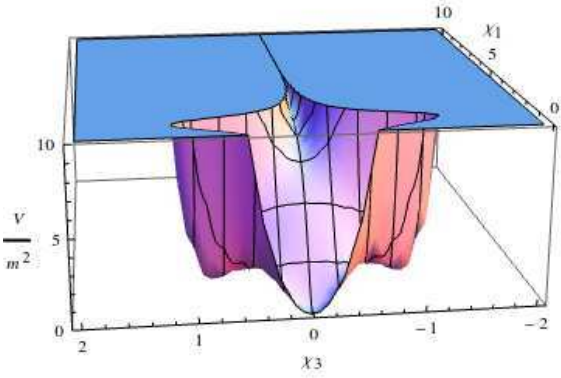}
  \caption{}
  \label{fig:sub1}
\end{subfigure}
\begin{subfigure}{.5\textwidth}
  \includegraphics[width=.8\linewidth]{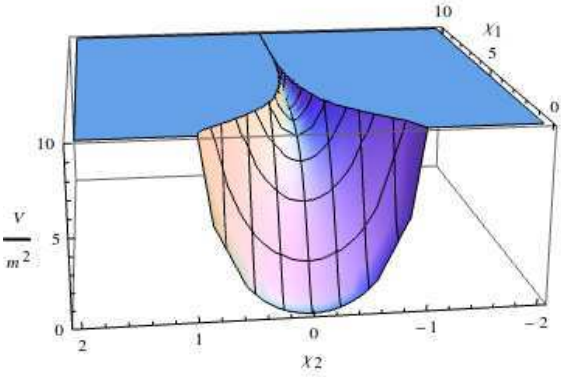}
  \caption{}
  \label{fig:sub2}
\end{subfigure}
\caption{The potential for the $SU(3)_C \times SU(2)_L \times SU(2)_R \times U(1)_{B-L}$
intermediate symmetry is shown. The inflation potential
is along $\chi_1$ direction. In Fig.1a we show V($\chi_1$, $\chi_2$ = 0, $\chi_3$)
and in Fig.1b V($\chi_1$, $\chi_2$, $\chi_3$ = 0). We see that potential
is flat along $\chi_1$ and confined along $\chi_2$ and $\chi_3$ respectively. \label{fig:test}}
\end{figure}

 {\bf Case I} : $a$ $\neq$ 0 and $p$ = $\omega$ = 0, $SU(3)_C \times SU(2)_L \times SU(2)_R \times U(1)_{B-L}$ symmetry.\\
 The kinetic and potential energy term are given by,
\begin{align} L_{K.E.}&=\frac{(1-a^2)(\partial_{\mu} p)^2+3(\partial_{\mu} a)^2+6(1-a^2)(\partial_{\mu} \omega)^2}{(1-a^2)^2},\nonumber\\
 V&=\frac{36 a^4 \lambda ^2+72 a^3 \lambda  m+36 a^2 m^2}{\left(1-a^2\right)^2}\,.\end{align}
 To get the canonical K.E. terms we need to redefine our fields in terms of
 new fields $\chi_1, \chi_2, \chi_3$,
 \be a= \tanh[\frac{\chi_1}{\sqrt{3}}], \,  p= \sech[\frac{\chi_1}{\sqrt{3}}] \chi_2, \,
  \omega=  \frac{1}{\sqrt{6}} \sech[\frac{\chi_1}{\sqrt{3}}] \chi_3 \,.\ee
  The potential V($\chi_1, \chi_2, \chi_3$) is flat along $\chi_1$ direction for $\chi_2$ = $\chi_3$ = 0 and is confined
  in the orthogonal ($\chi_2, \chi_3$) directions as shown in Fig. \ref{fig:test}.

\begin{figure}
  \includegraphics[width=.8\linewidth]{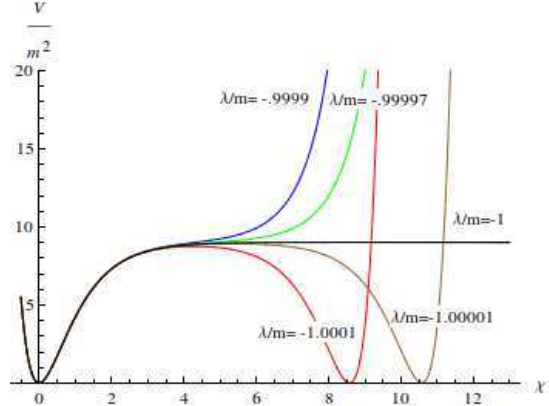}
  \caption{The potential $V/m^2$ for Case I for different chosen values of $\lambda/m$.}
  \label{vbymsq}
\end{figure}

 The potential V($\chi_1$) in the limit $\chi_2 = \chi_3$ = 0 is,
   \be V=\frac{36 \lambda^2 \tanh ^4\left[\frac{\chi_1 }{\sqrt{3}}\right]+72 m \lambda
   \tanh ^3\left[\frac{\chi_1 }{\sqrt{3}}\right]+36 m^2 \tanh
   ^2\left[\frac{\chi_1 }{\sqrt{3}}\right]}{\left(1-\tanh
   ^2\left[\frac{\chi_1 }{\sqrt{3}}\right]\right)^2}\ee
If we take $\lambda$= -$m$, this gives us the Starobinsky type of inflationary potential.
The potential in this specific case is,
\be V = 36 m^2(1-e^{-\frac{2 \chi_1}{\sqrt{3}}})^2 \,.\ee
This potential is shown in Fig. \ref{vbymsq} along with small deviations from the relation $\lambda$ = -$m$.
The slow roll parameters for this potential are given by,
\be \eta=-\frac{8 e^{\frac{-2 \chi_1 }{\sqrt{3}}} \left(1-2e^{\frac{-2 \chi_1 }{\sqrt{3}}}\right)}{3
   \left(1-e^{-\frac{2 \chi_1 }{\sqrt{3}}}\right)^2}; \quad  \epsilon= \frac{8 e^{-\frac{4 \chi_1 }{\sqrt{3}}}}{3 \left(1-e^{-\frac{2 \chi_1
   }{\sqrt{3}}}\right)^2}\,.\ee
   Inflation ends when $\eta$ $\approx$ 1, which corresponds to field value of $\chi_1^{end}$ $\approx$ 0.5.
   To have sufficient inflation which corresponds to $N_{e-folds}$=55 gives the initial field
   value of $\chi_1$ $\approx$ 4.35. The power spectrum for scalar perturbation $P_R$ is,
   \be P_R = \frac{V}{24\pi^2 \epsilon} = \frac{9 m^2 \sinh ^4\left(\frac{\chi_1 }{\sqrt{3}}\right)}{\pi ^2}\,.\ee
   The value of $P_R = (1.610 \pm 0.01) \times 10^{-9}$ given by Planck data \cite{PLANCK5} requires value of $m$ = $1.311 \times 10^{-6}$ in Planck units.
   The spectral index $n_s$ = .964 and tensor to scalar perturbation ratio r = .002 for $N_{e-folds}$=55.
   Varying $\lambda/m$ in the range (-1.0001 \textendash~ -0.9999) gives $n_s$
   in the range (0.92\textendash1.0) and $r$ in range (0.002 \textendash 0.008).\\

{\bf Case II}: $p$ $\neq$ 0 and $a$ = $\omega$ = 0, $SU(4)_C \times SU(2)_L \times SU(2)_R$ symmetry. \\
 The kinetic and potential energy term are given by,
 \begin{align}
 L_{K.E.}&=\frac{(\partial_{\mu} p)^2+3 (1-\frac{p^2}{3})(\partial_{\mu} a)^2
+6(1-\frac{p^2}{3})(\partial_{\mu} \omega)^2}{(1-\frac{p^2}{3})^2},\nonumber\\
 V&=\frac{4 m^2 p^2}{(1-\frac{p^2}{3})^2}
\,. \end{align}
 The fields transformation which make kinetic energy term canonical are,
\be p= \sqrt{3} \tanh[\frac{\chi_1}{\sqrt{3}}], \, a= \sech[\frac{\chi_1}{\sqrt{3}}]\frac{\chi_2}{\sqrt{3}} , \,
  \omega=   \sech [\frac{\chi_1}{\sqrt{3}}] \frac{\chi_3}{\sqrt{6}}\,. \ee
   Then the potential V($\chi_1$) in the limit $\chi_2 = \chi_3$ = 0 is,
   \be V=3 m^2 \sinh[\frac{2 \chi_1}{\sqrt{3}}]^2 \,.\ee
  This type of potential increases exponentially with $\chi_1$ and is too steep to obey the slow roll conditions.
The spectral index $n_s$ has negative values over a wide range of field value and hence doesn't satisfy the
inflationary constraints on scale invariance of scalar perturbations from observations.\\

{\bf Case III}: $\omega$ $\neq$ 0 and $p$ = $a$ = 0, $ SU(3)_C \times SU(2)_L \times U(1)_R \times U(1)_{B-L}$ symmetry. \\
The kinetic and potential energy term are given by,
\begin{align}L_{K.E.}&=\frac{(1-2\omega^2)(\partial_{\mu} p)^2+3 (1-2\omega^2)(\partial_{\mu} a)^2
+6(\partial_{\mu} \omega)^2}{(1-2\omega^2)^2},\nonumber\\
 V&=\frac{144 m^2 w^2+180 \lambda ^2 w^4}{\left(1-2 w^2\right)^2} \,.\end{align}
The fields transformation which makes kinetic energy term canonical are,
 \be \omega=  \frac{1}{\sqrt{2}} \tanh[\frac{\chi_1}{\sqrt{3}}], \, p= \sech[\frac{\chi_1}{\sqrt{3}}] \chi_2, \,
 a=\sech[\frac{\chi_1}{\sqrt{3}}] \frac{\chi_3}{\sqrt{3}} \,.\ee
   Then the potential V($\chi_1$) in the limit $\chi_2 = \chi_3$ = 0 is,
   \be V =72 m^2 \sinh[\frac{\chi_1}{\sqrt{3}}]^2(\cosh[\frac{\chi_1}{\sqrt{3}}]^2+\alpha \sinh[\frac{\chi_1}{\sqrt{3}}]^2 )\,. \ee
 Here $\alpha=5\lambda^2/8m^2$.  In this case for $\alpha \ge -1$ potential
 increases exponentially with $\chi_1$ and hence gives similar results as Case II.
 For $\alpha <$ -1 potential energy becomes negative for $\chi_1 \gtrsim $ 1 and grows with large values of $\chi_1$. Therefore this
 intermediate symmetry doesn't give successful inflation.\\

 {\bf Case IV}: If $p$ = $a$ = $\pm \omega$ $\neq$ 0, $ SU(5) \times U(1)$ symmetry. \\
In this case we take $p$ = $a$ = $\pm\omega$ = $x$, then the K.E. term and potential are given by,
\begin{align} L_{K.E.}&=\frac{90 (\partial_{\mu} x)^2}{\left(3-10 x^2\right)^2},\nonumber\\
 V&=\frac{184 m^2 x^2+1104 \lambda  m x^3+1656 \lambda ^2
   x^4}{\left(1-\frac{10 x^2}{3}\right)^2}\,.\end{align}
  The field redefinition $x = \sqrt{\frac{3}{10}} \tanh[\frac{\chi_1}{\sqrt{3}}]$ which makes kinetic energy term canonical gives the form of potential,
  \be V = 55.2 m^2(1-e^{-\frac{2 \chi_1}{\sqrt{3}}})^2\,,\ee
  for $\lambda=-\frac{1}{3}\sqrt{\frac{10}{3}}m$. This is a Starobinsky
  inflationary potential but with different relation among superpotential
  parameters $m$ and $\lambda$ in comparison to the Case I. In this case value of $m$ = 1.06 $\times 10^{-6}$ is required to satisfy the constraints
  from CMB observations. Small variations from the relation $\lambda = -\frac{1}{3}\sqrt{\frac{10}{3}}m$ gives the same types of deviations in the Starobinsky
  potential as shown in Fig. \ref{vbymsq}.

 At the end of inflation the inflaton $\chi_1$ can decay to scalar
 bosons which have a trilinear term with $\Phi$ in superpotential
 e.g. $ \Phi H (\gamma \Sigma+\bar \gamma \bar\Sigma)$.
 Then the $K_{\Sigma}^{\Sigma^*}$$|W_{\Sigma}|^2$ and $K_{\bar\Sigma}^{\bar\Sigma^*}$$|W_{\bar\Sigma}|^2$ type of terms gives,
 \be V \supset ((|\gamma|^2+|\bar\gamma|^2)|H|^2+|\gamma|^2|\Sigma|^2+|\bar\gamma|^2|\bar\Sigma|^2)|\sinh[\frac{\chi_1}{\sqrt{3}}]|^2\,. \ee
Near the origin $\sinh[\frac{\chi_1}{\sqrt{3}}]$ $\approx$ $\frac{\chi_1}{\sqrt{3}}$, so
\be V \supset ((|\gamma|^2+|\bar\gamma|^2)|H|^2+|\gamma|^2|\Sigma|^2+|\bar\gamma|^2|\bar\Sigma|^2)|\frac{\chi_1}{\sqrt{3}}|^2\,. \ee
In our case the  perturbative decay of inflaton to scalars is not efficient for
typical values of $\gamma$,$\bar\gamma$ $\sim$ O(.1-1.0) \cite{Aulakh:2013lxa}.
However inflaton $\chi_1$ can decay non-perturbativly to scalar bosons leading to preheating.
In \cite{preheating} the mechanism of preheating in broad resonance regime has been worked out.
There is another efficient way of preheating called ``instant preheating'' \cite{instpreheating}.
This mechanism is based upon the non-perturbative decay of inflaton to scalar bosons (in this case) when it is close to the minimum of the potential
(at $\chi_1$ = 0). The particles thus produced (having mass directly proportional to the instantaneous $vev$ of inflaton)
decay further when inflaton rolls uphill, to the modes which are not directly coupled to inflaton.
 This happens because at the time of their production, their mass is zero since $\chi_1$ = 0, but
as inflaton rolls back to its maximum value they become heavy so their decay width increases.
 In our case, every time inflaton crosses the origin it produces the $H, \Sigma$ and $\bar\Sigma$ . These
 decay further into the SM fermions and the right-handed neutrinos through Yukawa couplings.
 With this kind of chain reaction we can  have an efficient way to transfer the whole
energy of inflaton into  relativistic particles within  few
oscillations. This whole process leads to a radiation dominated universe with reheat temperature,
\be T_R \sim V_0^{1/4} \sim (m^2 \chi_1^2)^{1/4} \sim (10^{-18}M_{P}^4)^{1/4}\sim 10^{14} GeV \,.\ee 
 At the end of reheating, the universe has a finite temperature potential and after cooling from $T_R$ = $10^{14}$ GeV
  to temperature $<<$ $T_{R}$, we assume that universe settles to the minimum of potential corresponding to MSSM symmetry. The main requirement of
this new minimum is zero cosmological constant which can be achieved if the fields $a$, $p$, $\omega$, $\sigma(\bar\sigma)$ take values such that the scalar potential
 $V=|W_{\phi_i}|^2/{\Gamma^\prime}^2=0$ ( where
 $\Gamma^\prime=T+T^*-\frac{1}{3}(|p|^2+3|a|^2+6|\omega|^2+|\sigma|^2+|\bar\sigma|^2)$. The condition  $W_{\phi_i}$ = 0 required to have zero
 cosmological constant with broken SUSY (from the $vev$ of the
 moduli fields $T$ and $T^*$ ) in no-scale SUGRA is algebraically same as the condition for unbroken global supersymmetry in SUSY-SO(10)  \cite{abmsv}.
The field values  $a$, $p$, $\omega$, $\sigma(\bar\sigma)$ which give $W_{\phi_i}$ = 0 in SUSY SO(10) have been worked out in
\cite{abmsv} and are given by,
 \bea
a= \frac{m}{\lambda}\frac{x^2+2x-1}{1-x};\, p=\frac{m}{\lambda} \frac{x(5
x^2-1)}{(1-x)^2};\,\nonumber\\
\sigma\overline{\sigma}=\frac{2 m^2}{\eta \lambda} \frac{x (1-3 x)(1+x^2)}{\eta (1-x)^2};\,\, \omega = -\frac{m}{\lambda} x \label{mssm}\eea
where $x$ is the solution of following cubic equation,
 \bea 8 x^3-15 x^2+14 x -3 = -\frac{\lambda m_{\Sigma}}{\eta m} (1-x)^2 \,.\eea

 The soft SUSY breaking masses are proportional to the gravitino mass, which in no-scale SUGRA  models  with $V=0$ is
 given \cite{Lahanas,BasteroGil:1998te} by, \be m_{3/2}^2 = e^{G}=e^{K}|W|^2 \,.\label{GM}\ee
In our case visible sector also contributes to gravitino mass  as all the $vevs$ are in units of $m/\lambda$ so they can be
of O($M_{P}$) from the inflationary conditions. However visible sector contribution can be made zero or negligible with field values of $a$, $p$, $\omega$, $\sigma(\bar\sigma)$
given by Eq. (\ref{mssm}) and tuning  $|W|$ $\approx$ 0. In that case  only hidden sector and moduli fields determine the gravitino mass.

Also we need a pair of light Higgs doublets in MSSM. In the present scenario we have a $4\times4$
 mass matrix $\cal{H}$ of MSSM Higgs doublets \cite{aulgir1}. The form of mass matrix remains same as given in \cite{aulgir1} with an extra factor of $1/\Gamma'$,
 
 \begin{widetext}
  \be \begin{aligned}{\cal H}=\frac{1}{\Gamma^\prime}\begin{pmatrix} -m_H &
\bar{\gamma}\sqrt{3}(\omega-a) & -\gamma\sqrt{3}(\omega + a)&
-\bar{\gamma}\bar{\sigma}\cr
 -\bar{\gamma}\sqrt{3}(\omega+ a)& 0 & -(2m_{\Sigma} + 4\eta(a+ \omega))&0\cr
\gamma\sqrt{3}(\omega-a) & -(2m_{\Sigma} + 4\eta(a- \omega))&0 & -2\eta
\bar{\sigma}\sqrt{3}\cr
-\sigma\gamma & -2\eta\sigma\sqrt{3}&0 & -2m + 6\lambda(\omega-a) \end{pmatrix}
  \label{hmatrix} \,. \end{aligned}\ee \end{widetext}
 One out of the four Higgs doublets can be made light with the fine tuning condition of $Det \cal{H}$ = 0.
 For fixed values of $p$, $a$, $w$, $m$, $\lambda$, it can be solved for $m_H$ in terms of other free parameters of superpotential.
 For fixed real value of $x$ = -0.3471 from $|W|$ $\approx$ 0 in the cases of successful inflation, $m_H$ is given by,
\be m_H = \frac{-0.887 \bar\gamma \gamma}{\eta}~~(\text{case} ~I); \,\, m_H= \frac{-1.458\bar\gamma \gamma}{\eta} ~~(\text{case} ~IV)\,.\ee
For this $m_H$, one eigenvalue can be made light and the eigenvectors (left and right) corresponding to that eigenvalue can act as MSSM Higgs doublets.

\section{Conclusions}
In this work we show  that the Starobinsky model of inflation can be derived from no-scale
SUGRA SO(10) GUT for the specific intermediate symmetries of $SU(3)_C \times SU(2)_L \times SU(2)_R \times U(1)_{B-L}$,
$SU(5)\times U(1)$ and flipped $SU(5)\times U(1)$ gauge groups.
The other intermediate symmetries 
 $SU(4)_C \times SU(2)_L \times SU(2)_R$ or  $ SU(3)_C \times SU(2)_L \times U(1)_R \times U(1)_{B-L}$
 do not give the slow-roll potential required for inflation. In the course of symmetry breaking
 topological defects like monopoles and cosmic strings can form. The defects formed in the first stage of symmetry breaking
 SO(10) $\rightarrow$ intermediate scale takes place during
 inflation and will be diluted away. After reheating when intermediate symmetry breaks to MSSM topological defects may form  once again.
  The flipped $SU(5)\times U(1)$ and $SU(3)_C \times SU(2)_L \times SU(2)_R \times U(1)_{B-L}$ breaking down to MSSM produces 
  the  cosmic strings \cite{topodefects} type of defect
 which is acceptable. However $SU(5)\times U(1)$ gives rise to monopoles after inflation and this case therefore can 
 be ruled out from the consideration of topological defects
 in the cosmological evolution.
 The parameters of the SO(10) invariant superpotential are restricted by the requirement that the Starobinsky potential is obtained.
 These relations at the GUT scale can have testable consequences
in the particle spectrum at low energy.

\section{Acknowledgements}
We thank the anonymous referee for useful suggestions.

\end{document}